\documentclass[12pt]{article}
\begin{document}
\title{A Formula for the Mass Spectrum of Baryons and Mesons}
\author{B.G. Sidharth\\
Centre for Applicable Mathematics \& Computer Sciences\\
B.M. Birla Science Centre, Adarsh Nagar, Hyderabad - 500 063 (India)}
\date{}
\maketitle
\begin{abstract}
A mass formula for all the Baryons and Mesons is proposed. A
comparison with the actual masses shows that about $63 \%$ of the
errors are less than $1 \%$ while in about  $93 \%$ of the cases
errors are less than $2 \%$. In all cases the error is less than
$3 \%$ with the lone exception of $\omega (782)$ Meson in which
case the error is $3.6 \%$.A rationale for the proposed mass
formula is also touched upon.
\end{abstract}
\section{The Mass Formula}
We propose the formula
\begin{equation}
\mbox{mass} = m\left(n + \frac{1}{2}\right) \cdot 137 MeV\label{e1}
\end{equation}
for the mass of the elementary particles ($137 MeV$ in the pion mass, nearly). The results for Baryons and Mesons compared with the actual values from the Particle Data Group \cite{Hagiwara} are shown in Table1  and Table2.
\section{Comments}
i) For both Baryons and Mesons some $63 \%$ of the errors are less than $1 \%$ and some $93 \%$ errors are less than $2 \%$. In all cases the error is less than $3 \%$ with a single exception: the $\omega (782)$ Meson (error $3.6 \%$). In any case the actual errors are likely to be less than the figures indicated in the Table because of the uncertainty in the mass - these errors have been computed with respect to a value that is conventionally taken to be the mass, within the range of uncertainty.\\
When the mass formula value falls within the uncertainty range of the mass, the error has been shown as zero, though the error with respect to the conventional value has also been indicated.\\
In the case of Baryons, four stars against the particle indicate that its existence is certain and its properties are atleast fairly well explored. The lesser number of stars points to a lesser degree of certainty, a single star denoting the fact that the evidence for the existence of the particle is still poor. In the case of Mesons the star denotes the fact that these particles have been known from atleast the previous compilation.\\
ii) Interestingly the newly discovered $D_S$ particle \cite{Nature}, also fits in the table.\\
iii) It will be noticed that there is degeneracy on the one hand for certain particle masses, and also a large number of gaps for various values of $m$ and $n$ in (\ref{e1}), where particles and resonances have not yet been found.\\
iv) A rationale for the formula (\ref{e1}) can be found in the fact that, firstly the $\pi_0$ can be treated as the bound state of a positive and negatively charged particle, within the pion Compton wavelength, consistent with its decay into an electron positron pair as discussed in detail in \cite{MPLA1, CU}. Secondly the particles are treated as the energy levels of a tri-atomic type oscillator of two positive and one negative charge or the other way round (Cf.refs.\cite{MPLA1,CU,MPLA2,FFP2,CSF} for details). The different values that the positive integer $m$ takes comes from there being $m = 1,2,3,\cdots$, such oscillators constituting the particles, while $n$ defines the levels of the oscillators.

\begin{table}
\caption{Baryons}
\begin{tabular}{|c|c|c|} \hline
Particle and Mass & Mass from Formula &  Error \% \\ \hline
   $p(938)$ & $959$ &  $-2.23881,$ \\
   $n(939)$ & $959$ &  $-2.12993,$ \\
$P_{11} **** N(1440)$  & $1438$ & $(0.138889,)0$ \\
$D_{13}**** N(1520)$ & $1507$ &  $(0.855263,)$ \\
$S_{11}****N(1535)$ & $1540$  & $(-0.325733,)0$ \\
$S_{11}****N(1650)$ & $1644$  & $(0.363636,)0$ \\
$D_{15}****N(1675)$ & $1644$  & $1.85075,$ \\
$F_{15}****N(1680)$ & $1644$  & $2.14286,$ \\
$D_{13}***N(1700)$  & $1712$  & $(-0.705882,)0$ \\
$P_{11}***N(1710)$ & $1712$   & $(-0.116959,)0$ \\
$P_{13}****N(1720)$ & $1712$  & $(0.465116,)0$ \\
$P_{13}****N(1900)$ & $1918$  & $-0.947368,$ \\
$F_{17}**N(1990)$ & $1986$    & $0.201005,$ \\
$F-{15}**N(2000)$ & $1986$    & $0.7,$ \\
$D_{13}**N(2080)$ & $2055$    & $1.20192,$ \\
$S_{11}*N(2090)$ & $2123$     & $-1.57895,$ \\
$P_{11}*N(2100)$ & $2123$     & $(-1.09524,)$ \\
$G_{17}****N(2190)$ & $2123$  & $(3.05936,)0$ \\
$D_{15}**N(2200)$ & $2260$  & $-2.72727,$ \\
$H_{19}****N(2220)$ & $2260$  & $(-1.8018,)0$ \\
$G_{19}****N(2250)$ & $2260$  & $(-0.444444,)0$ \\
$I_{1;11}***N(2600)$ & $2603$ & $(-0.115385,)0$ \\
$K_{1;13}**N(2700)$ & $2671.5$ & $1.05556$ \\
$P_{33}****\Delta (1232)$ & $1233$ & $(-0.0811688,)0$ \\
$P_{33}***\Delta (1600)$ & $1575$  & $(1.5625,)0$ \\
$S_{31}****\Delta (1620)$ & $1644$ & $(-1.46148,)0$ \\
$D 33****\Delta (1700)$ & $1712$  & $(-0.705882,)0$ \\
$P 31*\Delta (1750)$ & $1781$  & $-1.77143,$ \\
$S_{31}**\Delta (1900)$ & $1918$  & $-0.947368,$ \\
$F 35****\Delta (1905)$ & $1918$  & $(-0.682415,)0$ \\
$P 31****\Delta (1910)$ & $1918$ & $(-0.418848,)0$ \\
$P 33***\Delta (1920)$ & $1918$  & $(0.104167,)0$ \\
$D 35***\Delta (1930)$ & $1918$  & $(0.621762,)0$ \\
$D 33*\Delta (1940)$ & $1918$  & $1.13402,$ \\
$F 37****\Delta (1950)$ & $1918$  & $1.64103,$ \\
$F 35**\Delta (2000)$ & $1986$  & $0.7,$ \\
$S_{31}*\Delta (2150)$ & $2123$ & $1.25581,$ \\
$G_{37}*\Delta (2200)$ & $2260$ & $-2.72727,$ \\
$H_{39}**\Delta (2300)$ & $2329$  & $-1.26087,$ \\
$D_{35}*\Delta (2350)$ & $2329$  & $0.893617,$ \\
$F_{37}*\Delta (2390)$ & $2397$  & $-0.292887,$ \\
$G_{39}**\Delta (2400)$ & $2397$ & $0.125,$ \\
\end{tabular}
\end{table}
\newpage

\begin{tabular}{|c|c|c|} \hline
Particle and Mass & Mass from Formula &  Error \% \\
$H_{3;11}****\Delta (2420)$ & $2397$  & $(0.950413,)0$ \\
$I_{3;13}**\Delta (2750)$ & $2740$  & $0.363636,$ \\
$K_{3;15}**\Delta (2950)$ & $2945.5$  & $0.152542,$ \\
              $\Lambda (1115)$ & $1096$  & $-4.33099,$ \\
$P_{01}****\Lambda (1600)$  & $1575.5$  &  $1.53125,$ \\
$S_{01}****\Lambda (1405)$ & $1431$ &  $-1.85053,$ \\
$D_{03}****\Lambda (1520)$ & $1507$ &  $0.855263,$ \\
$P 01***\Lambda (1600)$ & $1575$ &  $(1.5625,)0$ \\
$S 01****\Lambda (1670)$ & $1644$  &  $1.55689,$ \\
$D 03****\Lambda (1690)$ & $1712$  &  $-1.30178,$ \\
$S 01***\Lambda (1800)$ & $1781$  & $(1,05556,)0$ \\
$P 01***\Lambda (1810)$ & $1781$  & $(1.60221,)0$ \\
 $\Lambda (1820)$ & $1849$ & $(2.14286,)$ \\
$D 05****\Lambda (1830)$ & $1849$ & $-1.03825,$ \\
$P 03****\Lambda (1890)$ & $1918$  & $-1.48148,$ \\
$*\Lambda (2000)$ & $1986$  & $0.7,$      \\
$F 07*\Lambda (2020)$ & $2055$ & $-1.73267,$ \\
$G 07****\Lambda (2100)$ & $2123$  & $-1.09524,$ \\
$F 05***\Lambda (2110)$ & $2123$ & $(-0.616114,)0$ \\
$D 03*\Lambda (2325)$ & $2329$  & $-0.172043,$ \\
$H 09***\Lambda (2350)$ & $2329$  & $0.893617,$ \\
$**\Lambda (2585)$ & $2603$ & $0.309478,$ \\
$P 11****\Sigma +(118)$  & $1164$ & $2.10261,$ \\
$P 11****\Sigma 0 (119)$ & $1164$ & $2.34899,$ \\
$****\Sigma - (119)$ & $1164$  & $2.75689,$ \\
$P 13****\Sigma (1385)$  & $1385$ & $0,$       \\
$*\Sigma (1480)$  & $1438$ & $2.83784,$ \\
$**\Sigma (1560)$  & $1575$   & $-0.961538,$ \\
$D 13**\Sigma (1580)$  & $1575$ & $0.316456,$  \\
$S 11**\Sigma (1620)$  & $1644$ & $-1.48148,$  \\
$P 11***\Sigma (1660)$  & $1644$ & $(0.963855,)0$ \\
$D 13****\Sigma (1670)$  & $1644$ & $1.55689,$     \\
$**\Sigma (1690)$  & $1712$  & $-1.30178,$    \\
$S 11***\Sigma (1750)$  & $1781$ & $(-1.77143,)0$ \\
$P 11*\Sigma (1770)$  & $1781$  & $-0.621469,$ \\
$D 15****\Sigma (1775)$  & $1781$ & $(-0.338028,)0$ \\
$P 13*\Sigma (1840)$  & $1849$ & $-0.48913,$ \\
$P 11**\Sigma (1880)$  & $1849$   & $1.64894,$  \\
$F 15****\Sigma (1915)$  & $1918$ & $(-0.156658,)0$ \\
$D 13***\Sigma (1940)$  & $1918$ & $(1.13402,)0$ \\
$S 11*\Sigma (2000)$  & $1986$ & $0.7,$     \\
$F 17****\Sigma (2030)$  & $2055$ & $-1.23153,$ \\
$F 15*\Sigma (2070)$  & $2055$ & $0.724638,$ \\
$P 13**\Sigma (2080)$  & $2055$ & $1.20192,$ \\
$G 17*\Sigma (2100)$  & $2123$ & $-1.09524,$ \\
$***\Sigma (2250)$  & $2260$   & $(-0.444444,)0$ \\ \hline
\end{tabular}

\newpage

\begin{tabular}{|c|c|c|} \hline
Particle and Mass & Mass from Formula  & Error \% \\
$**\Sigma (2455)$  & $2466$ & $-0.448065,$ \\
$**\Sigma (2620)$  & $2603$ &  $0.648855,$ \\
$*\Sigma (3000)$  & $3014$  & $-0.466667,$ \\
$*\Sigma (3170)$  & $3151$  & $0.599369,$ \\
$P 11****\Xi 0,\Xi-(13)$ & $1301.5$  & $1.01156,$ \\
$****\Xi(1321)$  & $1301.5$  & $1.47615,$ \\
$P 13****\Xi (1530)$     & $1507$  & $1.50327,$ \\
$*\Xi (1620)$  & $1644$  & $-1.48148,$ \\
$***\Xi (1690)$     & $1712$  & $-1.30178,$ \\
$D 13***\Xi (1820)$     & $1849$ & $-1.59341,$ \\
$***\Xi (1950)$   & $1918$   & $1.64103,$  \\
$***\Xi (2030)$     & $2055$ & $-1.23153,$ \\
$*\Xi (2120)$     & $2123$   & $-0.141509,$ \\
$**\Xi (2250)$     & $2260$  & $-0.444444,$ \\
$**\Xi (2370)$     & $2397$  & $-1.13924,$ \\
$*\Xi (2500)$     & $2534$   & $-1.36,$    \\
$****\Omega -(1672)$ & $1644$  & $1.67464,$ \\
$***\Omega -(2250)$ & $2260$   & $(-0.444444,)0$\\
$**\Omega -(2380)$ & $2397$   & $-0.714286,$ \\
$**\Omega -(2470)$ & $2466$  & $0.161943,$ \\
$****\Lambda c+2285)$  & $2260$ & $1.09409,$ \\
$+***\Lambda c+(2593)$ & $2603$ & $-0.385654,$ \\
$+***\Lambda c+(2625)$ & $2603$ & $0.838095,$ \\
$+*\Lambda c+(2765)$ & $2740$  & $0.904159,$ \\
$+**\Lambda c+(2880)$ & $2877$  & $0.104167,$ \\
$****\Sigma c(2455)$   & $2466$  & $-0.448065,$ \\
$***\Sigma c(2520)$   & $2534$  & $-0.555556,$ \\
$\Xi c+(2466)$   & $2466$  & $0,$         \\
$***\Xi c0(2471)$     & $2466$  & $0.202347,$  \\
$***\Xi c+(2574)$     & $2577$  & $(-0.11655,)0$ \\
$***\Xi c0(2578)$     & $2577$  & $(0.0387898,)0$\\
$\Xi c(2645)$      & $2671$     & $-0.982987,$ \\
$***\Xi c(2790)$      & $2808$  & $-0.645161,$ \\
$***\Xi c(2815)$      & $2808$  & $0,248668,$ \\
$***\Omega c0(2697)$  & $2671$  & $0.964034,$ \\
$***\Lambda b0(5624)$ & $5617$  & $(0.124467,)0$ \\ \hline
\end{tabular}

\begin{table}
\caption{Mesons}
\begin{tabular}{|c|c|c|} \hline
Particle and mass & Mass From Formula &  Error \% \\ \hline
$*\pi^{\pm} (139)$ &  $137$ & $-1.43885$ \\
$*\pi^0 (135)$ & $137$ & $1.481481$  \\
$*\eta (547)$ & $548$ &  $0.182815$\\
$*f_0(600)$ & $616.5$ & $2.75$\\
$*\rho (770)$ & $753.5$ & $-2.14286$\\
$*\omega (782)$ & $753.5$ & $-3.6445$\\
$*\eta' (958)$ & $959$ & $0.104384$\\
$*f_0(980)$ & $ 959$ & $-2.14286$\\
$*a_0(980)$ & $ 959$ & $-2.14286$\\
$*\phi (1020)$ & $1027.5$ & $0.735294$\\
$*h_1 (1170)$ & $1164.5$ & $-0.47009$\\
$*b_1 (1235)$ & $1233$ & $-0.16194$ \\
$a_1 (1260)$ & $1233$ & $-2.14286$ \\
$f_2 (1270)$ & $1233$ &  $-2.91339$\\
$f_1 (1285)$ & $1301.5$ & $1.284047$ \\
$*\eta (1295)$ & $1301.5$ &  $0.501931$\\
$\pi (1300)$ & $1301.5$ &  $0.115385$\\
$a_2 (1320)$ & $1301.5$ & $-1.40152$\\
$*f_0 (1370)$ & $1370$ & $0$ \\
$h_1 (1380)$ & $1370$ &  $0.72464$\\
$\pi_1 (1400)$ & $1370$ & $-2.14286$ \\
$f_1 (1420)$ & $1438.5$ & $1.302817$ \\
$*\omega (1420)$ & $1438.5$ & $1.302817$ \\
$f_2 (1430)$ & $1438.5$ & $0.594406$ \\
$*\eta (1440)$ & $1438.5$ &  $-0.10417$\\
$*a_0 (1450)$ & $1438.5$ &  $-0.7931$\\
$*\rho (1450)$ & $1438.5$ & $-0.7931$ \\
$*f_0 (1500)$ & $1507$ & $0.466667$ \\
$f_1 (1510)$ & $1507$ &  $-0.19868$\\
$*f'_2 (1525)$ & $1507$ & $-1.18033$ \\
$f_2 (1565)$ & $1575.5$ & $0.670927$ \\
$h_1 (1595)$ & $1575.5$ &  $-1.22257$\\
$\pi_1 (1600)$ & $1575.5$ & $-1.53125$ \\
$\chi (1600)$ & $1575.5$ & $-1.53125$ \\
$a_1 (1640)$ & $1644$ & $0.243902$ \\
$f_2 (1640)$ & $1644$ &  $0.243902$\\ \hline
\end{tabular}
\end{table}
\newpage

\begin{tabular}{|c|c|c|} \hline
Particle and mass & Mass From Formula &  Error \% \\ \hline
$*\omega_3 (1670)$ & $1644$ &  $-1.55689$\\
$*\pi_2  (1670)$ & $1644$ & $-1.55689$ \\
$*\phi (1680)$ & $1712.5$ & $1.934524$ \\
$*\rho_3 (1690)$ & $1712.5$ & $1.331361$ \\
$*\rho (1700)$ & $1712.5$ & $0.735294$  \\
$a_2 (1700)$ & $1712.5$ &  $0.735294$\\
$f_0(1710)$ & $1712.5$ & $0.146199$ \\
$\eta (1760)$ & $1781$ &  $1.193182$\\
$*\pi (1800)$ & $ 1781$ & $-1.05556$ \\
$f_2(1810)$ & $1781$ & $-1.60221$ \\
$*\phi_3 (1850)$ & $1849.5$ & $-0.02703$ \\
$\eta_2 (1870)$ & $1849.5$ & $-1.09626$ \\
$\rho (1900)$ & $1918$ &  $0.947368$\\
$f_2 (1910)$ & $1918$ & $0.418848$ \\
$f_2 (1950)$ & $1918$ & $-1.64103$ \\
$\rho_3 (1990)$ & $1986.5$ & $-0.17588$ \\
$X (2000)$ & $1986.5$ & $-0.675$ \\
$f_2 (2010)$ & $1986.5$ & $-1.16915$ \\
$f_0 (2020)$ & $1986.5$ & $1.65842$ \\
$*a_4 (2040)$ & $2055$ & $0.735294$\\
$f_4 (2050)$ & $2055$ &  $0.243902$\\
$\pi_2 (2100)$ & $2123.5$ & $1.119048$\\
$f_0 (2100)$ & $2123.5$ &  $1.119048$\\
$f_2 (2150)$ & $2123.5$ &  $-1.23256$\\
$\rho_2 (2150)$ & $2123.5$ & $-1.23256$ \\
$f_0 (2200)$ & $2260$ & $2.75$\\
$f_J (2220)$ & $2260$ &  $1.824324$\\
$\eta (2225)$ & $2360$ & $1.595506$ \\
$\rho_3 (2250)$ & $2260$ & $0.466667$ \\
$*f_2 (2300)$ & $2329$ & $1.26087$ \\
$f_4 (2300)$ & $2329$ &  $1.26087$\\
$f_0 (2330)$ & $2329$ &  $-0.04292$ \\
$*f_2 (2340)$ & $2329$ & $-0.47009$  \\
$\rho_5 (2350)$ & $2329$ & $-0.89362$ \\ \hline
\end{tabular}

\newpage

\begin{tabular}{|c|c|c|} \hline
Particle and mass & Mass From Formula &  Error \% \\ \hline
$a_6 (2450)$ & $2466$ & $-0.89362$ \\
$f_6 (2510)$ & $2534.5$ & $0.976096$ \\
$*K^* (892)$ & $8905$ & $-0.16816$ \\
$*K_1(1270)$ & $1233/1395$ & $0$ \\
$*K_1(1400)$ & $1370$ & $-2.14286$\\
$*K^*(1410)$ & $1438.5$ & $2.021277$  \\
$*K^*_0(1430)$ & $1438.5$ & $0.594406$  \\
$*K^*_2(1430)$ & $1438.5$ & $0.594406$ \\
$K (1460)$ & $1438.5$ & $-1.4726$ \\
$K_2(1580)$ & $1575.5$ & $-0.28481$ \\
$K (1630)$ & $1644$ & $0.858896$ \\
$K_1 (1650)$ & $1644$ & $-0.36364$\\
$*K^* (1680)$ & $1712$ & $1.934524$ \\
$*K_2 (1770)$ & $1781$ & $0.621469$\\
$*K^*_3 (1780)$ & $1781$ & $0.05618$ \\
$*K_2 (1820)$ & $1849.5$ & $1.620879$ \\
$K (1830)$ & $1849.5$ & $1.065574$ \\
$K^*_0 (1950)$ & $1918$ & $-1.64103$ \\
$K^*_2 (1980)$ & $1986.5$ & $0.328283$ \\
$*K^*_4 (2045)$ & $2055$ & $0.488998$ \\
$K_2 (2250)$ & $2260.5$ & $0.466667$ \\
$K_3 (2320)$ & $2329$ &  $0.387931$\\
$K^*_5 (2380)$ & $2397$ &  $0.735294$\\
$K_4 (2500)$ & $2466$ & $-1.36$ \\
$K (3100)$ & $3082.5$ &  $-0.56452$\\
$*D^\pm (1869.3)$ & $1849.5$ & $-1.05922$  \\
$*D^\pm_0 (1968.5)$ & $1986.5$ & $0.914402$ \\
$*D^* (2007)$ & $01986.5$ & $-1.02143$  \\
$D^* (2010)$ & $\pm 1986.5$ & $-1.16915$ \\
$D_S (2317)$ & $2329$ & $0.51791$ \\
$*D_1 (2420)$ & $2397.5$ & $-0.92975$ \\
$D_1 (2420)$ & $\pm 2397.5$ & $-0.97067$ \\
$D^* (2460)$ & $02466$ & $0.243902$\\
$D^*(2460)$ & $\pm 2671.5$ &  $0.243902$ \\ \hline
\end{tabular}

\newpage

\begin{tabular}{|c|c|c|} \hline
Particle and mass & Mass From Formula &  Error \% \\ \hline
$D_{S1} (2536)$ & $\pm 2534.5$ & $-0.07885$  \\
$D_{SJ} (2573)$ & $2534.5$ & $-1.49631$ \\
$*B^{\pm} (5278)$ & $5274.5$ & $ -0.08524$  \\
$*B^0 (5279.4)$ & $5274.5$ & $-0.09281$ \\
$B_j(5732)$ & $5754$ & $-0.47009$ \\
$*B^0_S (5369.6)$ & $5343$ & $-0.49538$ \\
$B^*_{SJ} (5850)$ & $5822.5$ & $-0.47009$ \\
$*B^\pm_c (6400)$ & $6370.5$ & $0.4609$  \\
$*\eta c (1S) (2979)$ & $2945.5$ & $-1.12454$ \\
$*J/\psi (1S) (30968)$ & $3082.5$ & $-0.46402$ \\
$*\chi c_0 (1P) (3415.1)$ & $3425$ & $0.289889$ \\
$*\chi c_1 (1P) (3510.5)$ & $3493.5$ & $-0.48426$ \\
$*\chi c_2 (1P) (3556)$ & $3562$ & $0.168729$  \\
$*\psi (2S) (3685.9)$ & $3699$ &  $0.355408$\\
$*\psi (3770)$ & $3767.5$ & $(-0.06631)0$ \\
$*\psi (3836)$ & $3836$ & $0$ \\
$*\psi (4040)$ & $4041.5$ & $(0.037129)0$\\
$*\psi (4160)$ & $4178.5$ &  $(0.444712)0$\\\hline
\end{tabular}

\newpage

\begin{tabular}{|c|c|c|} \hline
Particle and mass & Mass From Formula &  Error \% \\ \hline
$*\gamma (1S) (9460.3)$ & $9453$  & $-0.07716$  \\
$\chi b_0 (1P) (9859.9)$ & $9864$ & $0.041583$ \\
$*\chi b_1 (1P) (9892.7)$ & $9864$ & $-0.29011$\\
$*\chi b_2 (1P) (9912.6)$ & $9864$ & $-0.49029$ \\
$*\gamma (2S) (10023)$ & $10001$ & $0.21949$\\
$*\chi b_0 (2P) (10232)$ & $10275$ & $0.42026$ \\
$*\chi b_1 (2P) (10255)$ & $10275$ & $0.1945027$ \\
$*\chi b_2 (2P) (10268)$ & $10275$ & $0.068173$ \\
$*\gamma (3S) (10355)$ & $10353$ & $-0.01931$ \\
$*\gamma (4S) (10580)$ & $10549$ & $-0.29301$ \\
$*\gamma (10860)$ & $10891.5$ & $0.290055$  \\
$*\gamma (11020)$ & $11028.5$ & $0.077132$ \\ \hline
\end{tabular}

\end{document}